\documentclass[aps, twocolumn,longbibliography,amsmath,amssymb,floatfix,pra,reprint,showpacs,footinbib,superscriptaddress]{revtex4-1}
\usepackage{epsfig,graphicx}
\usepackage{setspace}
\usepackage[english]{babel}
\usepackage{amsfonts}
\usepackage{amsmath}
\usepackage{latexsym}
\usepackage{graphics,bm}
\usepackage{natbib}
\usepackage{dcolumn}
\usepackage{bm}
\usepackage{rotating}
\usepackage{amsmath}
\usepackage{braket}
\usepackage{epstopdf}
\usepackage{color}

\usepackage{soul,xcolor}
\setstcolor{red}
\usepackage[T1]{fontenc}
\usepackage[applemac]{inputenc}
\usepackage{lmodern}
\usepackage{ae}
\usepackage{units}
\usepackage{color}
\usepackage{amsmath, amssymb, graphics, setspace}

\newcommand{\mathsym}[1]{{}}
\newcommand{\unicode}[1]{{}}

\usepackage{url}
\usepackage[colorlinks]{hyperref}
\hypersetup{%
        plainpages=true,
        breaklinks=true,
        hypertexnames=false,
        pageanchor=true,
        colorlinks=true,
        linkcolor={violet},
        citecolor={violet},
        urlcolor={violet},
        anchorcolor={black}
      }
\newcommand{\be}{\begin{equation}}
\newcommand{\ee}{\end{equation}}
\newcommand{\ba}{\begin{array}{lll}}
\newcommand{\ea}{\end{array}}

\newcommand{\lan}{\langle}
\newcommand{\ran}{\rangle}

\newcommand{\beg}{\begin{gather}}
\newcommand{\eeg}{\end{gather}}
\newcommand{\s}{\sigma}

\begin{document}
\title{Collective dynamics of inhomogeneously broadened emitters coupled to an optical cavity with narrow linewidth}
\author{Kamanasish Debnath}
\email[e-mail:]{kamanasish.debnath@phys.au.dk}
\affiliation{Department of Physics and Astronomy, Aarhus University, Ny Munkegade 120, DK-8000, Aarhus C, Denmark}
\author{Yuan Zhang}
\affiliation{Donostia International Physics Center, Paseo Manuel de Lardizabal 4, 20018 Donostia-San Sebastian (Gipuzkoa), Spain}
\author{Klaus M{\o}lmer}
\affiliation{Department of Physics and Astronomy, Aarhus University, Ny Munkegade 120, DK-8000, Aarhus C, Denmark}
\begin{abstract}
We study collective effects in an inhomogeneously broadened ensemble of two-level emitters coupled to an optical cavity with narrow linewidth. Using second order mean field theory we find that the emitters within a few times the cavity linewidth exhibit synchronous behaviour and undergo collective Rabi oscillations. Under proper conditions, the synchronized oscillations give rise to a modulated intracavity field which can excite emitters detuned by many linewidths from the cavity resonance. To study the synchronization in further detail, we simplify the model and consider two ensembles and study steady state properties when the emitters are subjected to an incoherent drive.
\end{abstract}
\maketitle

\section{Introduction}
There has been significant progress in the study of optical emitters in solids like NV centers in nanodiamonds~\cite{PhysRevLett.67.3420,Santori2010}, rare earth ions doped in crystals~\cite{PhysRevB.77.125111, HONG1998234}, quantum dots in nanoscale semiconductors~\cite{PhysRevLett.119.143601} etc. The scientific effort in the last decade has been dedicated equally towards investigating spectral properties of these systems~\cite{PhysRevLett.111.203601,Casabone2018} and also interfacing light with these systems using an optical cavity~\cite{Brachmann:16, Hunger_2010, PhysRevA.94.053835}. Advancements in this domain include superradiance with NV centers~\cite{Bradac2017, manual},  proposals for scalable quantum computing architectures ~\cite{Weber8513, OHLSSON200271, PhysRevA.69.022321, PhysRevA.75.012304}, Purcell enhanced decay~\cite{PhysRevApplied.6.054010, PhysRevLett.121.183603}, stable single photon sources~\cite{PhysRevLett.85.290, Brouri, PhysRevLett.121.133601} etc.

\begin{figure}
\centering
\includegraphics[width=0.48\textwidth]{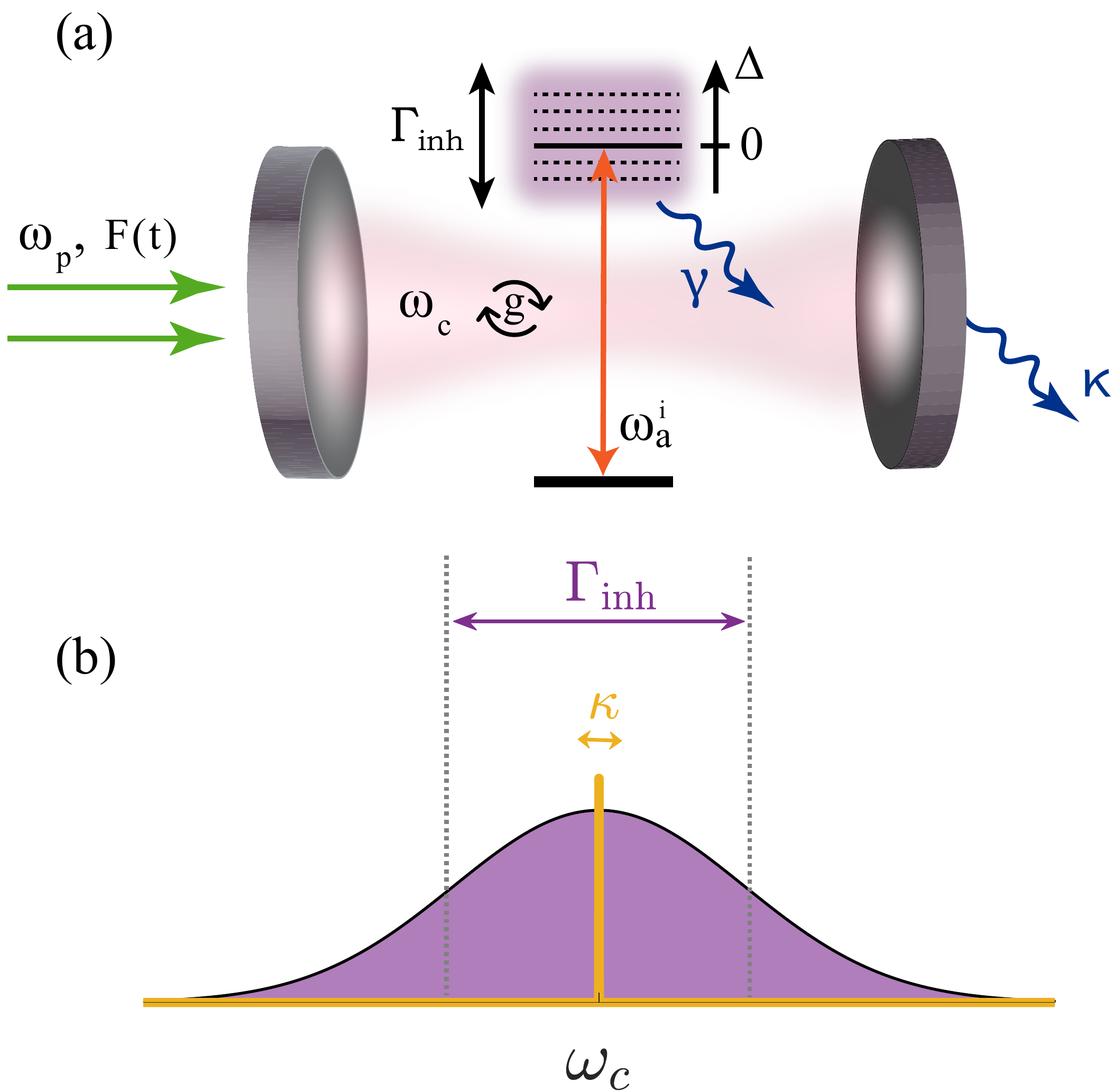}
\caption{Schematic of the system. (a) Ensemble of two-level emitters with inhomogeneously broadened excited level is coupled to an optical cavity of linewidth $\kappa$ and frequency $\omega_c$. The cavity is resonant with the central frequency $\omega^i_a$ of the inhomogeneous distribution. $\Delta$ corresponds to the detuning of the emitters due to the inhomogeneous broadening. The emitters are initialized by driving the cavity at resonance with an amplitude $F(t)$. (b) In this paper, the inhomogeneous linewidth $\Gamma_{\rm inh}$ is larger than the cavity linewidth $\kappa$.}
\label{F0}
\end{figure}

Due to interaction with the host environment, emitters in solids may suffer from vast intrinsic inhomogeneous broadening, which makes it difficult to observe coherent effects~\cite{GROSS1982301,PhysRevA.98.063815, PhysRevA.96.023863, PhysRevB.94.224510}. For systems like doped crystals or NV centers, the transition frequencies of all the emitters span across few GHz broad spectra~\cite{Casabone2018, PhysRevLett.111.203601, MACFARLANE20021}, while the linewidth of optical cavities ranges from few hundreds of kHz to few GHz. The vast inhomogeneity poses both a theoretical and an experimental challenge for the study of coherent light-matter interactions and for implementing these emitters for quantum computing applications.

Inhomogeneous systems, however, also offer the possibility for qubit encoding in frequency space~\cite{OHLSSON200271, PhysRevA.69.022321, PhysRevA.75.012304} for multi-mode quantum memories and they sometimes give rise to interesting physics like synchronization~\cite{PhysRevLett.121.063601}, quantum phase transition~\cite{doi:10.1063/1.4818403, Debnath_2017, arxivpreprint2,PhysRevA.100.013856, PhysRevA.100.013855}, many body localization~\cite{PhysRevLett.117.040601} etc.  Synchronization is an old classical phenomenon and has been reported in plethora of systems in diverse fields ranging from optomechanical arrays~\cite{PhysRevLett.111.073603, PhysRevLett.107.043603}, cold atomic ensembles~\cite{PhysRevLett.113.154101} in quantum optics~\cite{PhysRevLett.113.154101} to fireflies and pendula in non-linear dynamics~\cite{strogatz_2015}.

In this paper, we study the synchronization dynamics of an inhomogeneously broadened ensemble of two-level emitters coupled to an optical cavity, subjected to a coherent drive. There has been studies, both theoretically~\cite{PhysRevA.84.063810, PhysRevLett.95.243602, PhysRevA.53.2711} and experimentally~\cite{PhysRevX.8.021036, PhysRevX.7.031002} on inhomogeneous systems coupled to a cavity, but a majority of them concentrated on the case where the cavity linewidth was larger or similar to the inhomogeneous linewidth of the emitters. In our case, we consider the opposite regime where the inhomogeneous broadening is much larger than the linewidth of the cavity, at least by one order of magnitude. Systems which fall in this regime include NV centers and different species of rare earth ions doped in host crystals. For controllable test studies, ultracold atoms in a high finesse optical cavity may be spectrally broadened by an inhomogeneous magnetic field~\cite{PhysRevX.8.021036}. In addition, the Doppler broadening naturally induces inhomogeneous effects in atomic gases~\cite{PhysRevA.57.R3169, arxivpreprint}.

We find that a large group of emitters with frequencies within several linewidths of the cavity resonance display synchronized behaviour and perform collective Rabi oscillations. The frequency of these oscillations depends on the number of emitters within the cavity linewidth and the light-matter coupling. In the presence of a large number of emitters, this collective behaviour generates a strongly modulated field inside the cavity which, under the right conditions, is capable of exciting a select group of emitters, resonant with the corresponding sideband to the bare cavity mode.

Finally, we investigate the synchronization effect in the case when the emitters are subjected to an incoherent drive. We simplify the model to two ensembles and study the coherences between the emitters and verify that the synchronization persists under the incoherent drive.

The paper is arranged as follows. In Sec.~\ref{II} we describe the model followed by a brief description of the numerical approach used for the simulations. We present results of our calculations in Sec.~\ref{III}, where we first consider the complete model and explain the observed results with semi-analytical arguments and then we study a simplified toy model, which confirms the synchronous behaviour by investigating the steady state properties. We summarize our main findings in Sec.~\ref{IV}.

\section{Model \label{II}}
We consider a collection of two level emitters whose inhomogeneous broadening $\Gamma_{\rm inh}$ is larger than the linewidth $\kappa$ of the optical cavity to which they are coupled (i.e. $\Gamma_{\rm inh} > \kappa$) as shown in Fig.~\ref{F0}. We consider a general model which can be applied to a variety of inhomogeneous systems. Defining $\omega^i_a$ as the transition frequency and $g_i$ as the coupling of the i$^{\rm th}$ emitter with the cavity mode, the Hamiltonian of the system of $N$ emitters can be written as ($\hbar=1$ throughout the paper)
\begin{equation}
H= \delta_c a^{\dagger}a + \sum_{i=1}^{N}\Big[\frac{\delta^i_a}{2} \sigma^{i}_z + g_i(a\sigma^{i}_+ + a^{\dagger}\sigma^{i}_-)\Big] + F(t)(a + a^{\dagger}),
\label{E1}
\end{equation}
where $F(t)$ is the time dependent strength of the coherent pump with frequency $\omega_p$. $a, a^{\dagger}$ are the annihilation and creation operator of the optical mode in the frame rotating with the pump frequency (i.e. $\delta_c= (\omega_c - \omega_p)$ and $\delta^i_a= (\omega^i_a - \omega_p)$), respectively, and they follow the usual commutation relations $[a, a^{\dagger}] =1$. $\sigma^i_z, \sigma^i_+, \sigma^i_-$ are the Pauli operators. The dynamics of the system can be described by the master equation 
\begin{equation}
\begin{array}{lll}
\partial\rho/\partial t&=& -i[H,\rho] + \kappa \mathcal{D}[a]\rho + \gamma\sum_{i=1}^{N}\mathcal{D}[\sigma^i_-]\rho \\
&+& \Gamma\sum_{i=1}^{N}\mathcal{D}[\sigma^i_z]\rho,
\end{array}
\label{E2}
\end{equation}
where $\gamma$, $\Gamma$ are the decay and dephasing rate of the individual emitters respectively. The superoperator is defined as: $\mathcal{D}[\mathcal{O}]\rho= \mathcal{O}\rho \mathcal{O}^{\dagger} - \frac{1}{2}\{\mathcal{O}^{\dagger}\mathcal{O}, \rho\}$, for any operator $\mathcal{O}$.

Throughout this paper we shall use parameters which are relevant to systems like Eu$^{3+}$ ions doped in Y$_2$O$_3$ crystals~\cite{Casabone2018}. The homogeneous linewidth of the emitters in Eu$^{3+}$:Y$_2$O$_3$ varies from few kHz~\cite{Casabone2018} to few MHz~\cite{HONG1998234} and typically depends linearly on the temperature~\cite{PhysRevLett.111.203601}. Due to computational advantage, we initially choose a high value of the decay rate $\gamma/2\pi = 2$ MHz. We choose the coupling to be $g/2\pi=1.6$ kHz and an optical cavity of linewidth $\kappa/2\pi= 160$ kHz. The inhomogeneous broadening reported in the experiments~\cite{Casabone2018} was $22$ GHz, however, we initially restrict our calculations to $\Gamma_{\rm inh}/2\pi= 3.75$ MHz, following a Gaussian distribution with mean $\delta_c$ and standard deviation $10\kappa$. In Appendix~\ref{A2}, we extend the study to a peaked distribution, where the number of emitters with detuning $\Delta= (\delta_c - \delta^i_a)$ is proportional to $1/|\Delta|$ with $\Delta_{\rm max}/2\pi= 0.1$ GHz as the maximum value of the detuning. The parameters are also compatible with the optical transition for NV centers~\cite{Bradac2017}, however, their complex level scheme due to phononic sidebands at room temperature~\cite{PhysRevLett.110.243602} may not be well described by a model using two-level systems.

We consider $N= 10^8$ emitters in our model. Solving Eq.~\eqref{E2} for the full density matrix of the many emitters is impossible, and we treat the model by discretizing the inhomogeneous distribution into $k= 220$ frequency classes, following a Gaussian distribution with mean $\delta_c$ and standard deviation $10\kappa$. Further, we assume each frequency class to have identical and non distinguishable emitters and we employ a second order mean field theory to describe the dynamics of the system. In addition, we consider identical coupling between single emitters and the cavity. This approximation is justified in the regime of weak excitation, where the ensemble collective excitation dynamics in each frequency class depends only on the rms coupling strength. From Eq.~\eqref{E2}, we calculate the equation of motion for the mean intracavity photon number $\langle a^{\dagger}a\rangle$, which couples to the correlations between the cavity and the emitters $\langle a \sigma^{i}_+\rangle$, which in turn couple to the correlation between the emitters in different frequency classes i.e. $\langle \sigma^{k}_+\sigma^{k'}_-\rangle$. These second-order correlations, in turn, couple to third order correlations resulting in a hierarchy of equations. In order to truncate this hierarchy, we approximate any third-order correlation by products of lower-order correlations: $\langle ABC \rangle \approx \lan AB\ran\lan C\ran + \lan BC\ran\lan A\ran + \lan AC\ran\lan B\ran - 2\lan A\ran\lan B\ran \lan C\ran$. The resulting set of equations is reported in Appendix~\ref{A1}. We integrate these equations to study the dynamics of the system.

One can also employ a first order mean field theory with the approximation $\lan AB\ran \approx \lan A\ran \lan B \ran$ to describe the dynamics. However, such an approximation neglects the correlations between emitters mediated via the cavity mode, which is a crucial component of this study. The maximum practical number of frequency classes that can be simulated with our code parallelized on GPUs is approximately $220$. Our aim to investigate the dynamics of ensembles with frequencies both inside and outside the cavity linewidth restricts the FWHM ($\Gamma_{\rm inh}$) of our Gaussian distribution to $3.75$ MHz.

\section{Results \label{III}}
\begin{figure}
\centering
\includegraphics[width=0.5\textwidth]{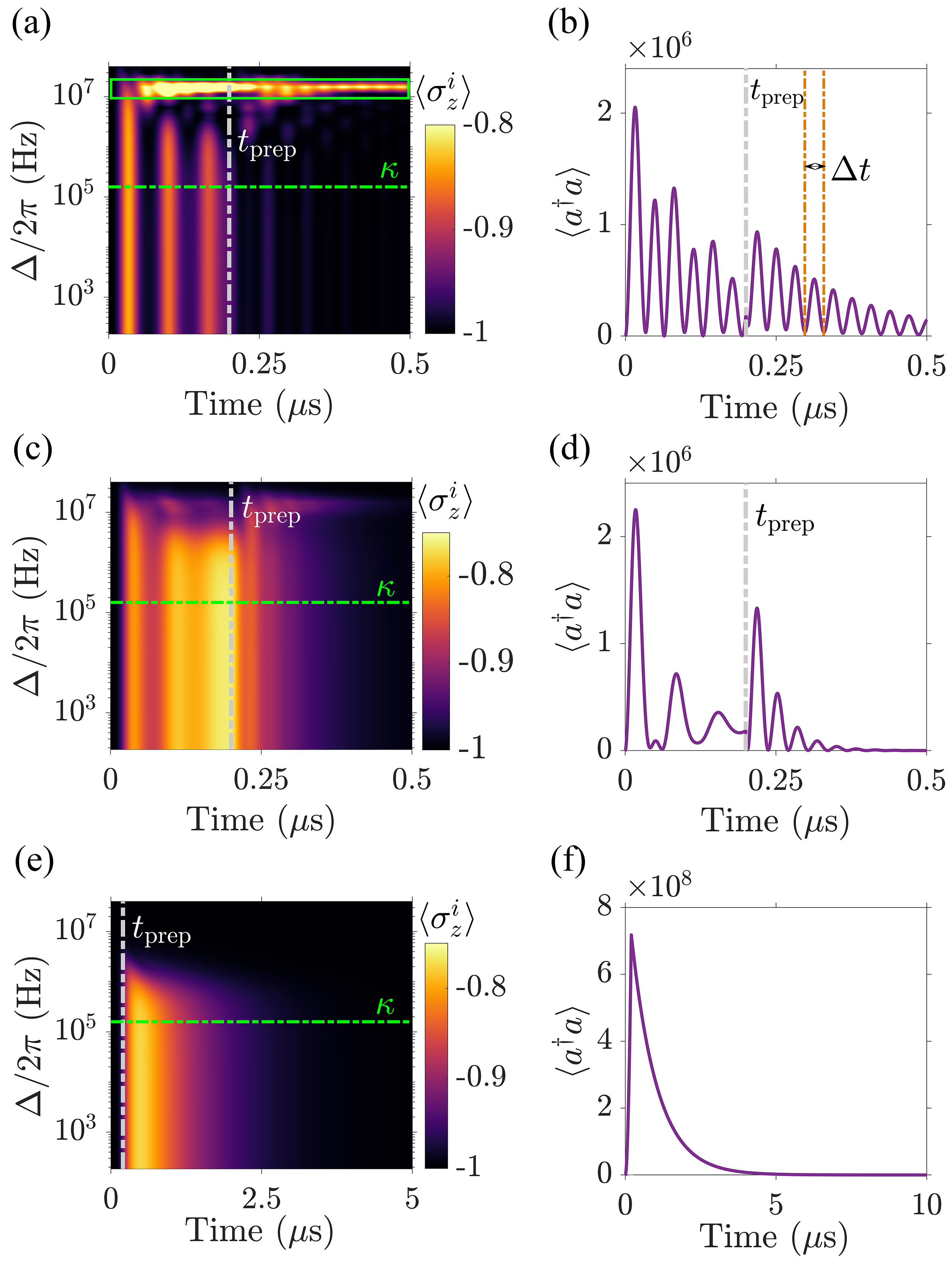}
\caption{Population inversion $\lan\sigma^i_z\ran$ for emitters in frequency classes with positive detuning, $\Delta$ as a function of time for (a) $g/2\pi= 1.6$ kHz and $\Gamma=0$, (c) $g/2\pi= 1.6$ kHz and $\Gamma/2\pi= 3.18$ MHz, (e) $g/2\pi= 16$ Hz and $\Gamma=0$.
(b), (d), (f) are the corresponding variation of the intracavity photon number associated with the population inversion shown in the left panels.}
\label{F2}
\end{figure}
In this section, we discuss the collective response of the inhomogeneously broadened ensemble upon driving. First, we consider the case when the cavity is subjected to a coherent drive and we investigate the transient dynamics. We apply a strong square pulse of duration $t_{\rm prep}= 0.2$ $\mu$s and strength $F= 2\pi \times 3 \times 10^7$ Hz resonant with the cavity frequency to excite the cavity and the emitters from their ground states, and subsequently, we study the exchange of energy between the emitters and the cavity followed by their decay. In the next subsection we simplify the model and consider the case when the emitters are subjected to an incoherent drive leading to a steady state emission.

\subsection{Transient Synchronization \label{A}}
We initially consider the case when the single emitter-cavity coupling is $g/2\pi= 1.6$ kHz. Fig.~\ref{F2}(a) shows the dynamics of single emitters in different frequency classes. The results are the same for negative as for positive $\Delta$. The cavity linewidth is marked with the horizontal dot-dashed green  line in the figure and the vertical dot-dashed white line marks the end of the excitation pulse $t_{\rm prep}$.  For $t\leq t_{\rm prep}$, as evidenced by the homogeneous bright and dark regions well beyond the cavity linewidth $\kappa$ in Fig.~\ref{F2}(a), the emitters exhibit synchronous behaviour. By synchronization, we explicitly mean that the emitters exhibit identical population and coherence dynamics, irrespective of their detuning. After the drive is turned off at $t= 0.2$ $\mu$s, the energy remaining in the emitters oscillates back and forth to the cavity mode until it is gradually lost. Around $\Delta/2\pi\approx 10^7$ Hz, we also observe a peculiar, non synchronized  excitation of the emitters (shown within the green rectangle) in a region detuned far beyond the dot-dashed green line. The origin of this peculiar long lived excitations of the emitters shall be explained later in the text.

In Fig.~\ref{F2}(b), we plot the intracavity photon number for the same parameters as in panel (a). During the driving pulse we observe a complex oscillatory evolution of both the excitation of the emitters and the photon number. This can be explained as follows. The collective emitter-cavity coupling is given by $g\sqrt{\mathcal{N}}$, where $\mathcal{N}$ is the number of emitters collectively coupled to the cavity mode. When this coupling exceeds the cavity decay rate $\kappa$, the photons emitted after $t_{\rm prep}$ can be coherently reabsorbed multiple times by the emitters before leaking out of the cavity. This leads to the damped Rabi oscillations after the preparation pulse is switched off. This behaviour was recently observed with trapped ultracold atoms in an optical cavity~\cite{PhysRevX.8.021036}.

Let us denote the frequency of the Rabi oscillations by $\Omega$. The decay rate of these oscillations is determined jointly by $\kappa$, $\gamma$ and $\Gamma_c$, where $\Gamma_c= g^2\mathcal{N}/\kappa$ is the Purcell rate of the emitters. The frequency $\Omega$ can be extracted from the period of the oscillations $\Delta t$ (shown with dot-dashed orange lines) as $\Omega/2\pi\approx 1.5 \times 10^7$ Hz. This acts as an intensity modulation of the intra-cavity field and thus provides sidebands to its carrier frequency which can excite emitters with detunings $\Delta= \pm\Omega = \pm 2\pi\times10^7$ Hz as observed in Fig.~\ref{F2}(a). The Purcell enhanced decay of this excitation is suppressed due to the detuning from the cavity resonance.

It is interesting to note that the oscillation frequency of $\lan\sigma^i_z\ran$ and $\lan a^{\dagger}a\ran$ in Fig.~\ref{F2} (a) and (b) differ by a factor of two. The difference in frequency is due to the evolution of the excitation amplitude $\lan\sigma_-\ran$ around a finite offset value, while the field amplitude evolves around zero. These coherent excitation amplitudes indeed oscillate at the same frequency $\Omega$, but the resulting number of excitations for $\lan\sigma_z\ran$ and $\lan a^{\dagger}a\ran$, proportional to $|\lan\sigma_-\ran|^2$ and $|\lan a\ran|^2$, show features at frequency $\Omega$ (due to the offset) and $2\Omega$ respectively.

The synchronization between the emitters is still preserved in the presence of dephasing as shown in Fig.~\ref{F2}(c), while for a weaker coupling, Fig.~\ref{F2}(d) shows the absence of collective Rabi oscillations and hence of any resonant sideband excitation.

\begin{figure}
\centering
\includegraphics[width=0.48\textwidth]{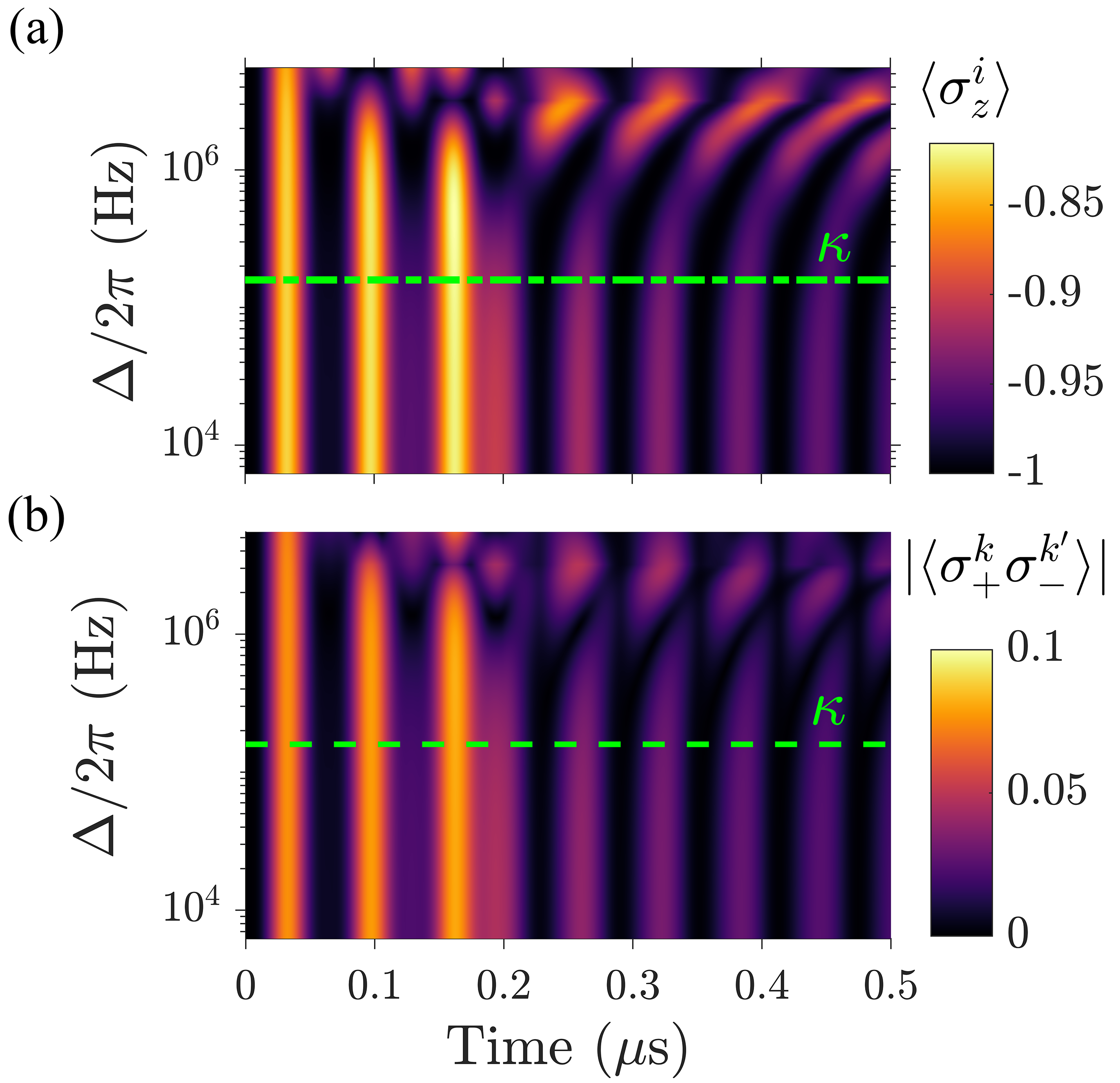}
\caption{(a) Population inversion $\lan\sigma^i_z\ran$ for emitters in different frequency classes as a function of time for $\gamma/2\pi= 1.6$ kHz. Other parameters are same as in Fig.~\ref{F2}(a).
(b) Correlation between emitters in different frequency classes, defined by $|\langle\sigma^k_+\sigma^{k'}_-\rangle|$, as a function of $\Delta$, the detuning of the $k'$th frequency class in our calculations while we assume $k=1$ corresponding to the frequency class which is resonant with $\omega_c$.}
\label{F3}
\end{figure}

In Fig.~\ref{F2}(e) and (f), the coupling is reduced to $g/2\pi = 16$ Hz and $\Gamma=0$. The Purcell enhanced decay rate, given by $\Gamma_c$, in this case, gets reduced by four orders of magnitude. The reduction in the effective Rabi frequency by two orders of magnitude explains the absence of Rabi oscillations in the emitter excitations in Fig.~\ref{F2}(e) and the intracavity photon number in Fig.~\ref{F2}(f).

The results presented in Fig.~\ref{F2} correspond to $\gamma/2\pi= 2$ MHz, which is comparable to the critical detuning $\Delta$, where the synchronization tends to break. This raises a natural question of whether the linewidth broadening due to large $\gamma$ is solely responsible for the synchronization effect. In Fig.~\ref{F3}(a), we show the transient dynamics for $\gamma= 1.6$ kHz ($\gamma\ll\kappa$), while keeping all other parameters the same as in Fig.~\ref{F2}(a). It is evident from the plot that the synchronization persists even for small $\gamma$. Due to reduction in the decay rate of the emitters, the observed Rabi oscillations are more prominent and long lived compared to Fig.~\ref{F2}(a).

The motivation to employ second order mean field theory in this study is to take into account the correlations, which are neglected in a first order treatment. The coherences between the emitters belonging to different frequency classes, $\lan\sigma^{k}_+\sigma^{k'}_-\ran$ are mediated by the cavity field, since the emitters do not interact directly. In Fig.~\ref{F3}(b), we plot $|\lan\sigma^{1}_+\sigma^{k'}_-\ran|$ as a function of time, where $k'$ is varied from $1$ to $220$. Clearly, the synchronization is supported by the presence of high, non-zero cross correlations between emitters in different frequency classes. The correlations decrease around the same detuning $\Delta$ region where the synchronous Rabi oscillation dynamics cease to occur.

In Appendix~\ref{A2}, we consider a frequency distribution, $N_k(\Delta) \propto 1/|\Delta|$, in the range $-10^8$ Hz $\le \Delta/2\pi \le$ $10^8$ Hz. Our simulations confirm that the features observed here with a Gaussian distribution remain valid even for a different distribution. In addition, we also verify in Fig~\ref{F5} (c) in Appendix~\ref{A2} that the sideband detuning $\Omega$ increases with the number of emitters, leading to a shift of the far detuned excitations.

We have shown that a large group of emitters exhibit identical dynamics and non zero cross correlation between emitters in different frequency classes, which we refer to as synchronization in analogy with the Huygens classical experiment with pendula. A multitude of factors play a role in determining the synchronized behaviour, namely the detuning $\Delta$, cavity linewidth $\kappa$, the Purcell rate $\Gamma_c$ and the emitter linewidth $\gamma$. In an attempt to quantify the most important factors, we simplify the model into two ensembles and investigate how the interplay between $\Delta, \kappa, \gamma$ and $\Gamma_c$ determine synchronization in a steady state scenario in the next subsection.

\begin{figure*}
\centering
\includegraphics[width=1\textwidth]{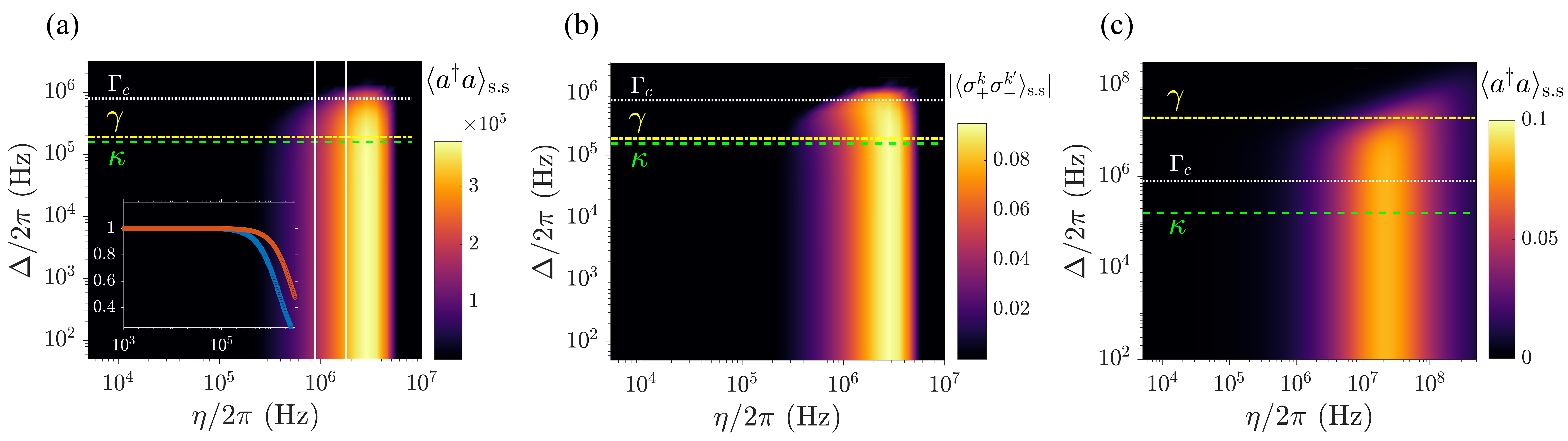}
\caption{Steady state (a) photon number $\lan a^{\dagger}a\ran_{\rm s.s}$ and
(b) coherence between emitters in different ensembles, $|\lan\sigma^k_+\sigma^{k'}_-\rangle_{\rm s.s}|$ as a function of incoherent pump and detuning when $\Gamma_c > \gamma$.
Inset in panel (a) shows the ratio of coherences between emitters in different and same frequency classes, $\Sigma$, for different values of $\eta$ (shown with white vertical line) as a function of detuning (in X axis).
(c) $\lan a^{\dagger}a\ran_{\rm s.s}$ as a function of detuning and pump intensity when $\gamma>\Gamma_c$.
Parameters chosen are: (a),(b) $\gamma/2\pi= 0.2$ MHz, $\kappa/2\pi= 160$ kHz (c) $\gamma/2\pi= 20$ MHz, $\kappa/2\pi= 160$ kHz.}
\label{F4}
\end{figure*}

\subsection{Steady State Synchronization \label{B}}

The previous subsection dealt with transient synchronization, when an inhomogeneous ensemble of emitters is subjected to a coherent drive. One may also be interested in the case when the emitters are incoherently pumped, which leads to a steady state emission (lasing)~\cite{PhysRevA.88.063825,PhysRevB.93.205148,PhysRevLett.102.163601,PhysRevA.99.033802}. If the synchronization effect holds also for the driven-dissipative case, one may observe narrow linewidth lasing with an inhomogeneous ensemble because synchronization causes emission at a single, common frequency. We introduce incoherent excitation of the emitters with rate $\eta$ and the master Eq.~\eqref{E2} gets modified as $\partial \rho_{\rm new}/\partial t = \partial \rho/\partial t + \eta \sum^N_{i=1} \mathcal{L}[\sigma^i_+]\rho$. This leads to few additional terms in the equations of motion. To reach more clear conclusions on the role of synchronization for the spectrum of the steady state radiation field, we simplify the model into two ensembles $A$ and $B$, with $N= 10^5$ emitters and transition frequency $\omega_A$ and $\omega_B$ respectively. A similar model has been studied to explore synchronization in the $\gamma \ll \kappa$ regime in~\cite{PhysRevLett.113.154101}.

In Fig.~\ref{F4}(a), we plot the steady state intracavity photon number $\lan a^{\dagger}a\ran_{\rm s.s}$ as a function of detuning and pump intensity. The steady state of a driven-dissipative system emerges due to the balance between the drive ($\eta$) and the decay rates ($\kappa$, $\gamma$ and $\Gamma_c$). As the pumping rate $\eta$ overcomes the decay rates, the steady state intracavity photon number starts increasing and beyond a limit, it drops due to saturation effects. These regimes are described in the literature as subradiance, superradiance and saturation depending on the number of steady state photons per emitter~\cite{Hollandarxiv, PhysRevA.98.063837}.

As shown in Fig.~\ref{F4}(a), $\langle a^{\dagger}a\rangle_{\rm s.s}$ drops abruptly as the detuning $\Delta$ surpasses the Purcell rate, given by $\Gamma_c\equiv g^2 N/\kappa$ (shown with horizontal dotted white line). Note that in this case, $\Gamma_c$ is larger than both the emitter linewidth $\gamma$ (dot-dashed yellow line) and the cavity linewidth $\kappa$ (dotted green line). This was also reported in a different study~\cite{PhysRevLett.113.154101}, but operating in the regime where $\kappa \gg \gamma, \Gamma_{\rm inh}$.

In Fig.~\ref{F4}(b), we plot the coherence between the emitters in different ensembles, $\lan\sigma^k_+\sigma^{k'}_-\ran$, where $k\neq k'$. As evident from the plot, once the pumping rate overcomes the decay rates, the system exhibits high values of coherences in the regime where the system is synchronized. As $\Delta$ exceeds $\Gamma_c$, $\lan\sigma^k_+\sigma^{k'}_-\ran$ drops abruptly. Further, we define $\Sigma$ as the ratio of $\langle\sigma^k_+\sigma^{k'}_-\rangle$ and the coherence between different emitters in the same frequency class, $\langle\sigma^{k,i}_+\sigma^{k,i'}_-\rangle$
 and plot $\Sigma$ as a function of $\Delta$ for different values of $\eta$ (shown with vertical white lines) in the inset of Fig.~\ref{F4}(a). The ratio of the coherences confirms that the two ensembles are synchronized ($\Sigma=1$) until $\Delta$ increases beyond $\Gamma_c$.

For $\gamma >\Gamma_c,\kappa$, Fig.~\ref{F4}(c) shows that synchronous behaviour of the spins persists for detunings as large as $\gamma$ and may be ascribed to the linewidth broadening of the emitters.

\section{Conclusion\label{IV}}
To summarize, we have studied the interaction of an inhomogeneously broadened ensemble of emitters coupled to an optical cavity with linewidth smaller than the inhomogeneous broadening. The parameters chosen are consistent with recent experiments with systems with rare earth ions doped in host crystals. We showed that a large group of emitters can show synchronous behaviour irrespective of their detunings. When the effective coupling between the emitters and the cavity is stronger than the cavity decay rate, the Rabi oscillations lead to the generation of a strong intracavity field with frequency sidebands at $\pm\Omega$. These sidebands can in turn interact resonantly with the emitters in the wings of the inhomogeneous distribution.

In an attempt to quantify the factors which determine synchronization, we simplified the model into two ensembles and investigated the steady state properties when the emitters are subjected to an incoherent drive. We observed a sudden drop in the cross coherences between the emitters at the onset of the non synchronous regime when the detuning exceeded the Purcell enhanced rate or the emitter linewidth, whichever is dominating. We suspect that a similar criterion for synchronization holds in the transient regime of the coherently driven system, while it is an open question how many emitters contribute effectively to the collective coupling and hence determine the value of $\Gamma_c$.

\section{Acknowledgements \label{V}}
K.D and K.M acknowledge support from the European Union's Horizon 2020 research and innovation program (No. 712721, NanoQtech).

\appendix
\section{Second order mean field equations \label{A1}}
The Hamiltonian of $N$ emitters coupled to an optical cavity is given by Eq.\eqref{E1}. Since solving the master equation exactly is impossible, we approximate the system by discretizing the inhomogeneous spectrum into $k= 220$ ensembles with $N_k$ identical emitters in each ensemble. The Hamiltonian can thus be written as
\begin{align}
H =& \delta_c a^{\dagger}a + \sum_{k=1}^{220}\sum_{i=1}^{N_k}\Big[\frac{\delta^{k,i}_a}{2} \sigma^{k,i}_z + g_{k,i}(a\sigma^{k,i}_+ + a^{\dagger}\sigma^{k,i}_-)\Big] \nonumber \\
\qquad &+ F(t)(a + a^{\dagger}).
\label{AB1}
\end{align}
We assume that all the emitters in each frequency class $k$ are identical with same detuning $\delta^{k,i}_a$ and coupling strength $g_i$. Hence, for simplicity, we drop the superscript $i$ and retain only $k$, which labels the frequency class (or ensemble) $k$ to which it belongs.
The mean field equation for any operator $\mathcal{O}$ can be calculated using
\begin{equation}
\frac{\partial}{\partial t}\mathcal{\lan O} \ran = {\rm Trace}\{\dot{\rho}\mathcal{O}\},
\end{equation}
where $\partial\rho/\partial t$ is given by Eq.~\eqref{E2} plus an additional term due to the incoherent drive explained before. We begin with the equation for the mean intracavity photon number
\begin{align}
\frac{\partial}{\partial t}\lan a^{\dagger}a\ran =& -2\sum^{220}_{k=1}g_kN_k {\rm Im} \lan a\sigma^k_+\ran - iF(t)(\lan a\ran - \lan a^{\dagger}\ran)\nonumber \\
\qquad &- \kappa\lan a^{\dagger}a\rangle,
\end{align}
where $\sigma^{k}_{\pm}$ represent the Pauli operators describing the emitters in the $k^{th}$ frequency class. The equation for the mean intracavity photon number couples to the mean cavity field $\lan a \ran$, which is given by
\begin{align}
\frac{\partial}{\partial t}\lan a \ran =& -i\tilde{\omega}_c \lan a \ran - i\sum^{220}_{k=1}g_k N_k\lan\sigma^k_-\ran - iF(t),
\end{align}
where $\tilde{\omega}_c = \delta_c - i(\kappa/2)$.
In addition, we would also require the equation for $\lan a^2\ran$ when we truncate the third order correlations later, which is given by
\begin{align}
\frac{\partial}{\partial t}\lan a^2 \ran =& -2i\tilde{\omega}_c \lan a^2 \ran - 2i\sum^{220}_{k=1}g_kN_k(\lan a\sigma^k_-\ran + \lan a\sigma^k_+\ran) \nonumber \\
\qquad &-2iF(t)\lan a\ran.
\end{align}
The interaction of the cavity field with all the frequency classes results in the summation over all ensembles from $k=1$ to $220$. The above equations couple to $\lan \sigma^k_-\ran$ and also to the correlations between the emitters and the cavity mode, namely $\lan a\sigma^k_+\ran$ and $\lan a\sigma^k_-\ran$. The equations for these correlations and their further dependencies are given by
\begin{align}
\frac{\partial}{\partial t}\lan \sigma^k_-\ran =&  -i\tilde{\omega}_k + ig_k\lan a\sigma^k_z\ran, \\
\frac{\partial}{\partial t}\lan \sigma^k_z\ran =&  4g_k{\rm Im}\lan a\s^k_+\ran - \gamma(1+\lan\s^k_z\ran)+ \eta(1-\lan\s^k_z\ran), \\
\frac{\partial}{\partial t}\lan a \s^k_z\ran =& -i\tilde{\omega}_c \lan a \s^k_z\ran - ig_k[\lan \s^k_z\ran + (N_k - 1)\lan\s^{k,i}_-\s^{k,i'}_z\ran] \nonumber \\
\qquad & - i\sum_{k\neq k^{'}}g_{k^{'}} N_{k^{'}} \lan\s^k_z\s^{k^{'}}_-\ran - 2ig_k[\lan a^2\s^k_+\ran \nonumber \\
\qquad & - \lan aa^{\dagger}\s^k_-\ran] - \gamma(\lan a\ran + \lan a\s^k_z\ran) + \eta(\lan a\ran - \lan a\s^k_z\ran) \nonumber \\
\qquad & - iF(t)\lan \s^k_z\rangle,
\end{align}

\begin{align}
\frac{\partial}{\partial t}\lan a\s^k_- \ran =& -i(\tilde{\omega}_c + \tilde{\omega}_k)\lan a\s^k_-\ran - i\sum_{k\neq k^{'}}g_{k^{'}}N_{k^{'}}\lan \s^k_-\s^{k^{'}}_-\ran \nonumber \\
\qquad &-ig_k[(N_k-1)\lan\s^{k,i}_-\s^{k,i'}_-\ran - \lan a^2\s^k_z\ran]- iF(t)\lan \s^k_-\ran, \\
\frac{\partial}{\partial t}\lan a\s^k_+ \ran =& i(\tilde{\omega}^*_k-\tilde{\omega}_c)\lan a\s^k_+ \ran - i(g_k/2)(1-\lan\s^k_z\ran) \nonumber \\
\qquad & -ig_k(N_k-1)\lan\s^{k,i}_-\s^{k,i'}_+\ran \nonumber \\
\qquad & - i\sum_{k\neq k^{'}}g_{k^{'}}N_{k^{'}}\lan\s^k_+\s^{k^{'}}_-\ran - ig_k\lan aa^{\dagger}\s^k_z\ran \nonumber \\
\qquad & - iF(t)\lan \s^k_+\ran,
\end{align}
where $\tilde{\omega}_k= \delta^k_a - i[(\gamma + \eta)/2 + \Gamma]$. The third order correlations in the above equations like $\lan a^2\sigma^k_+\ran$ and $\lan a a^{\dagger}\sigma^k_-\ran$ are approximated as the product of lower order correlations using the formula: $\langle ABC \rangle \approx \lan AB\ran\lan C\ran + \lan BC\ran\lan A\ran + \lan AC\ran\lan B\ran - 2\lan A\ran\lan B\ran \lan C\ran$ (e.g. $\lan a^2\sigma^k_+\ran = \lan a^2\ran\lan\sigma^k_+\ran + \lan a\sigma^k_+\ran\lan a\ran + \lan a\sigma^k_+\ran\lan a\ran - 2\lan a\ran \lan a\ran \lan\sigma^k_+\ran$). The terms with prefactor $(N_k - 1)$ in the above equations correspond to the correlations between different emitters within the same frequency class. Since each emitter interacts with all other emitters, hence the prefactor $(N_k - 1)$. In addition, each emitter in $k^{\rm th}$ frequency class also interacts with $N_k^{'}$ emitters in $k^{' \rm th}$ frequency class mediated by the cavity mode. Such interactions lead to the terms with prefactor $N_{k^{'}}$. The equations for the correlations between the emitters in the same frequency class are given by
\begin{align}
\frac{\partial}{\partial t}\lan \s^{k,i}_-\s^{k,i'}_z \ran =& -i\tilde{\omega}_k\lan\s^{k,i}_-\s^{k,i'}_z\ran  + ig_k\lan a\s^{k,i}_z\s^{k,i'}_z\ran \nonumber \\
\qquad &- 2ig_k(\lan a\s^{k,i}_-\s^{k,i'}_+\ran - \lan a^{\dagger}\s^{k,i}_-\s^{k,i'}_-\ran) \nonumber\\
\qquad &- \gamma(\lan\s^k_-\ran + \lan\s^{k,i}_z\s^{k,i'}_z\ran)+ \eta(\lan\s^k_-\ran \nonumber \\
\qquad &- \lan\s^{k,i}_z\s^{k,i'}_z\ran), \\
\frac{\partial}{\partial t}\lan \s^{k,i}_-\s^{k,i'}_+ \ran =& 2{\rm Im}(\tilde{\omega}_k)\lan \s^{k,i}_-\s^{k,i'}_+ \ran + 2g_k{\rm Im}\lan a^{\dagger}\s^{k,i}_-\s^{k,i'}_z\ran,
\end{align}
\begin{align}
\frac{\partial}{\partial t}\lan\s^{k,i}_-\s^{k,i'}_-\ran =& -2i\tilde{\omega}_k\lan\s^{k,i}_-\s^{k,i'}_-\ran + 2ig_k\lan a \s^{k,i}_-\s^{k,i'}_z\rangle, \\
\frac{\partial}{\partial t}\lan \s^{k,i}_z\s^{k,i'}_z \ran =& 8g_k{\rm Im}\lan \s^{k,i}_+\s^{k,i'}_z\ran - 2\gamma(\lan\s^k_z\ran + \lan \s^{k,i}_z\s^{k,i'}_z\ran) \nonumber \\
\qquad & + 2\eta(\lan\s^k_z\ran - \lan \s^{k,i}_z\s^{k,i'}_z\ran).
\end{align}
Finally, the equations governing the correlations between emitters in different frequency classes $k$ and $k^{'}$ are given by
\begin{align}
\frac{\partial}{\partial t}\lan \s^k_-\s^{k^{'}}_- \ran =& -i(\tilde{\omega}_k + \tilde{\omega}_{k^{'}})\lan \s^k_-\s^{k^{'}}_- \ran + ig_k\lan a\s^k_z\s^{k^{'}}_-\ran \nonumber \\
\qquad &+ ig_{k^{'}}\lan a\s^k_-\s^{k^{'}}_z\ran, \\
\frac{\partial}{\partial t}\lan \s^{k}_z\s^{k^{'}}_z \ran =& 4g_k{\rm Im}\lan a\s^k_+\s^{k^{'}}_z\ran + 4g_{k^{'}}{\rm Im}\lan a\s^{k^{'}}_+\s^k_z\ran \nonumber \\
\qquad & -\gamma(\lan\s^{k^{'}}_z\ran + \lan \s^k_z\s^{k^{'}}_z\ran) + \eta(\lan\s^{k^{'}}_z\ran \nonumber \\
\qquad & - \lan \s^k_z\s^{k^{'}}_z\ran )  - \gamma(\lan\s^k_z\ran + \lan \s^k_z\s^{k^{'}}_z\ran) + \eta(\lan\s^k_z\ran  \nonumber \\
\qquad & - \lan \s^k_z\s^{k^{'}}_z\ran ),
\end{align}
\begin{align}
\frac{\partial}{\partial t} \lan \s^k_z\s^{k^{'}}_- \ran =& -i\tilde{\omega}_{k^{'}}\lan \s^k_z\s^{k^{'}}_- \ran + ig_{k^{'}}\lan a\s^k_z\s^{k^{'}}_z\ran \nonumber \nonumber \\
\qquad & - 2ig_k(\lan a\s^k_+\s^{k^{'}}_-\ran - \lan a^{\dagger}\s^k_-\s^{k^{'}}_-\ran)  \nonumber \\
\qquad & - \gamma(\lan\s^{k^{'}}_-\ran + \lan\s^k_z\s^{k^{'}}_-\ran) + \eta(\lan\s^{k^{'}}_-\ran \nonumber \\
\qquad & - \lan\s^k_z\s^{k^{'}}_-\ran), \\
\frac{\partial}{\partial t}\lan \s^k_+\s^{k^{'}}_- \ran =& -i(\tilde{\omega}_{k^{'}} - \tilde{\omega}^*_k)\lan \s^k_+\s^{k^{'}}_- \ran + ig_{k^{'}}\lan a\s^k_+\s^{k^{'}}_z\ran \nonumber \\
\qquad &- ig_k\lan a^{\dagger}\s^k_z\s^{k^{'}}_-\rangle.
\end{align}
The third order correlations are approximated as described before. For simplicity, we consider ensembles with homogeneous coupling, i.e. $g_i=g_j=g$, which is justified in the weak excitation regime investigated here. For an ensemble with $k$ frequency classes, the total number of equations are $3 + 9k + 4k^2$. This includes $3$ equations for the cavity mode, $9k$ equations for each ensemble and their correlations and $4k^2$ equations for the correlations between different ensembles mediated by the cavity field.

\section{$1/|\Delta|$ distributed frequency classes \label{A2}}

In the main text, we modelled the inhomogeneous broadening $\Gamma_{\rm inh}$ as a normal distribution with a FWHM of $\Gamma_{\rm inh}= 3.75$ MHz. Here, we consider frequency classes distributed as $N_k(\Delta)\propto 1/|\Delta|$, where $N_k$ is the number of emitters in each frequency class $k$. We initialize the system by driving the cavity with the same pulse as in the main text, and study the dynamics of the single emitter in each frequency classes.

\begin{figure*}
\centering
\includegraphics[width=1\textwidth]{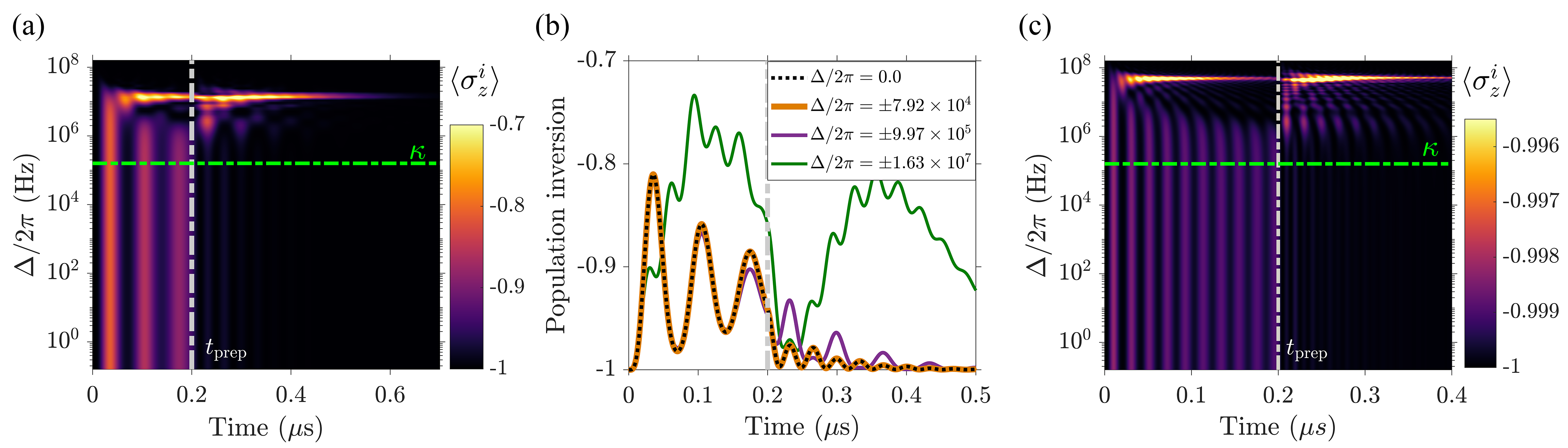}
\caption{(a) Population inversion $\lan\sigma^i_z\ran$ of emitters in all the frequency classes with positive detuning $\Delta$ as a function of time. (b) Population inversion $\lan\sigma^i_z\ran$ for emitters in selected frequency classes. (c) $\lan\sigma^i_z\ran$ of emitters in all the frequency classes with positive detuning for $N= 10^9$ and other parameters remaining the same as in panel (a).}
\label{F5}
\end{figure*}

In Fig.~\ref{F5}(a), we plot the excitation of the single emitters in different frequency classes as a function of time and we find that the synchronization effect persists. The excitation of the far detuned emitters due to the intensity modulation is also prominent for the emitters with $\Delta/2\pi\approx 1.5 \times 10^7$ Hz. The synchronization effect is also evident in Fig.~\ref{F5}(b), where we have shown the population inversion for selected horizontal sections from Fig.~\ref{F5}(a). In Fig.~\ref{F5}(c), we plot the population inversion $\lan\sigma^i_z\ran$ for $N= 10^9$, keeping all other parameters the same as in (a).

\end{document}